\documentclass[twocolumn,eqsecnum,aps]{revtex4}
\usepackage[dvips]{epsfig,color}
\usepackage{graphicx}
\usepackage{amsmath,amsfonts,amsbsy,amssymb}

\def\PRL{Phys. Rev. Lett. }
\def\PRB{Phys. Rev. B }

\def\be{\begin{eqnarray}}
\def\ee{\end{eqnarray}}

\def\bbox#1{\boldsymbol#1}

\def\spin{\text{spin}}

\def\upA{\uparrow}
\def\dnA{\downarrow}

\def\spin{\text{spin}}
\def\ppin{\text{ppin}}

\def\be{\begin{equation}}
\def\ee{\end{equation}}
\begin{document}
%%%%%%%%%%%%%%%%%%%%%%%%%%%%%%%%
\title{
 SU(4) Coherent  Effects to the Canted Antiferromagnetic Phase \\
in Bilayer Quantum Hall Systems at $\nu$=2}
%%%%%%%%%%%%%%%%%%%%%%%%%%
\author{ K. Hasebe}
\affiliation{ Department of Physics, Sogang University, Seoul, 100-611 Korea}
\begin{abstract}
\vspace*{5mm}
In bilayer quantum Hall (BLQH)  systems at $\nu$=2, 
three  different kinds of ground states are expected to be realized,
i.e. a spin polarized phase (spin phase), a pseudospin polarized phase
(ppin phase) and  a canted antiferromagnetic phase (C-phase).
An SU(4) scheme gives a powerful tool to investigate BLQH systems 
which have not only the spin SU(2)  but also the  layer (pseudospin)
SU(2) degrees of freedom. 
In this paper, we discuss an origin of the C-phase in the SU(4) context
 and investigate SU(4) coherent effects to it.
 We show  peculiar operators in the SU(4) group
which do not exist in SU$_{\text{spin}}$(2)$\otimes$SU$_{\text{ppin}}$(2)
group play a key role to its realization. It is also pointed out that not 
only spins but also pseudospins are ``canted'' in the C-phase. 
\end{abstract}

\maketitle
\section{Introduction}
Recently  bilayer quantum Hall (BLQH) systems have much attention 
due to their peculiar phenomena stemmed from  many isospin  degrees of 
freedom.
 BLQH systems have not only the spin SU(2) 
 but also the layer (pseudospin or ppin) SU(2) degrees of freedom.
The interlayer interactions bring 
 many  interesting phenomena  in 
 BLQH systems \cite{d5}.
 For instance, skyrmion excitations arise not only in a spin space 
 but also in a pseudospin space 
\cite{Ezawa97B,Ezawa99L}. 
 A Josephson-like  tunneling current across the layers 
 has been observed \cite{e1}, which is predicted a decade ago \cite{j2,j1}. 
 An SU(4) formalism  naturally incorporates the spin SU(2) 
 and pseudospin (ppin) SU(2) degrees of freedom. 
 To use such a larger group, one can easily treat 
 various ground states and excitations from a unified point of view
\cite{Hasebecondmat0110072,EzawaPRB65}. 

 In BLQH systems at $\nu$=2, 
three different kinds of phases are expected to be realized.
 One is a fully spin polarized ferromagnetic phase, which we call the spin phase. 
Second one is a spin singlet phase, which we call the ppin
phase because it is also a fully pseudospin polarized one. 
They have been observed experimentally \cite{Pelle97,Sawada98L,Sawada99R}.
The last one is a canted antiferromagnetic phase (C-phase),
 which is  predicted \cite{c1,c2,c3,c4,MacDonaldPhysRevB99,c5}
 between the spin and ppin phase in a phase diagram.
The C-phase is an exotic phase, where spins on the front layer and the 
 back layer   exhibit  ferromagnetic properties  
 in the perpendicular direction and simultaneously
   antiferromagnetic  properties 
 to the two dimensional layer    directions.

The spin and ppin  phases can be understood as different ground states from an 
SU(4) unified point of view. 
Excitations on two phases are determined by 
the breaking patterns of the SU(4) symmetry \cite{Hasebecondmat0110072}.  
Thus the SU(4) formalism has played  essential roles to understand  
physical properties of these phases.
However, SU(4) analysis on the C-phase   has not been accomplished at all.
The aim of this paper is to investigate the C-phase in the SU(4) context.

 Demler and Das Sarma analyzed the 
C-phase  with the use of an effective boson theory \cite{c4}.
Their Hamiltonian has not included
 the interlayer exchange interaction.
MacDonald et. al. also investigated the system based on the Hartree
 Fock method \cite{MacDonaldPhysRevB99}.
However isospin exchange interactions  have been dropped in the approximation.
Both  interlayer and isospin interactions are important in BLQH systems
 because they bring rich physics as stated above.

We improve their arguments  
with the use of more realistic SU(4) Hamiltonian, which includes the
interlayer  and SU(4) isopin exchange interactions  fully \cite{Hasebecondmat0110072,EzawaPRB65}. 
It is shown  
that  the SU(4) coherence  disturbs
 the C-phase realization.
 Peculiar SU(4) operators 
which do not exist in SU$_{\text{spin}}$(2)$\otimes$SU$_{\text{ppin}}$(2)
act as an  order parameter of the C-phase.
We refer them as $U$-operators, which
 are essential  for the C-phase realization.

In Section \ref{SecLandaHamil} we  review the derivation of 
the microscopic Landau-site
Hamiltonian  \cite{Hasebecondmat0110072,EzawaPRB65}
for the BLQH system, which is an anisotropic SU(4)
nonlinear sigma model in a continuum limit.
 In Section \ref{SecGrounState} we make a group
theoretical study of isospin states in  BLQH systems.
 At $\nu =2$ they belong to a $\boldsymbol{6}$-dimensional
 representation of SU(4). 
They are referred as Schwinger bosons \cite{SB}. 
With the use of the  boson picture, we construct an effective boson
theory just like \cite{c4}.
Classifying
them with respect to the subgroup SU$_{\text{spin}}$(2)$\otimes $SU$_{\text{%
ppin}}$(2), we introduce three different kinds of ground states, i.e. 
the spin phase, the   ppin phase 
 and the C-phase.               
It is shown that not only  spins but also ppins are 
``canted'' in the C-phase.
We discuss a crucial role of $U$-operators for the realization of  the C-phase.
In Section \ref{SecSU(4)Cohe}
 we investigate   SU(4) coherent effects to   the C-phase.
 The SU(4) coherence  reduces the 
C-phase region and  disturbs its realization.
We  give physical interpretations of this effect.
Section \ref{SecSummary} is devoted to summary and discussion.

\section{Landau-Site Hamiltonian}

\label{SecLandaHamil}

We analyze electrons in the lowest Landau level\ in BLQH systems. One Landau
site contains four electron states distinguished by the SU(4) isospin index $%
\sigma =\text{f}\!\uparrow ,\text{f}\!\downarrow ,\text{b}\!\uparrow ,\text{b%
}\!\downarrow $.

The Coulomb interaction is given by 
%%%%%%%%%%%%%%%%%%%%%%%%%%%%%%%%%%%%%%%%%%%%%%%%%%%%
\begin{equation}
H_{C}=\sum_{\alpha,\beta =\text{f},\text{b}}{\frac{1}{2}}\int d^{2}xd^{2}y
V_{\alpha \beta}(\boldsymbol{x}-\boldsymbol{y})\rho_{\alpha}(\boldsymbol{x})
\rho_{\beta} (\boldsymbol{y}),
\end{equation}
%%%%%%%%%%%%%%%%%%%%%%%%%%%%%%%%%%%%%%%%%%%%%%%%%%%%%
where $V_{\text{f}\text{f}}=V_{\text{b}\text{b}}={e^2}/4\pi \varepsilon r$
is the intralayer Coulomb interaction, while $V_{\text{f}\text{b}}=V_{\text{b%
}\text{f}}={e^2}/4\pi \varepsilon \sqrt {r^2+d^2}$ is the interlayer Coulomb
interaction with the interlayer separation $d$.
 The Coulomb interaction is
decomposed into two terms, $H_{C}=H_{C}^{+}+H_{C}^{-}$, with  
%%%%%%%%%%%%%%%%%%%%%%%%%%%%%%%%%%%%%%%%%%%%%%%%%%%%%%%
\begin{subequations}
\label{BLCouloPM}
\begin{align}
H_{C}^{+}&= {\frac{1}{2}}\int d^{2}xd^{2}y V_{+}(\boldsymbol{x}-\boldsymbol{y})\rho (%
\boldsymbol{x})\rho (\boldsymbol{y}) ,  \label{BLCouloP} \\
H_{C}^{-}&= {\frac{1}{2}}\int d^{2}xd^{2}y V_{-}(\boldsymbol{x}-\boldsymbol{y}%
)\Delta \rho (\boldsymbol{x})\Delta \rho (\boldsymbol{y}),  \label{BBLCouloM}
\end{align}
\end{subequations}
%%%%%%%%%%%%%%%%%%%%%%%%%%%%%%%%%%%%%%%%%%%%%%%%%%%%%%%
where $V_{\pm }={\frac{1}{2}}(V_{\text{f}\text{f}}\pm V_{\text{f}\text{b}})$%
; $\rho (\boldsymbol{x})$ is the total density , while $%
 \Delta \rho (\boldsymbol{x})$ is the density difference between the
front and back layers. 
$H_{C}^{+}$ respects the SU(4) symmetry, while $H_{C}^{-}$ does not.

By using the von-Neumann formalism \cite{vonNeumann55,EzawaPRB65},
 the exchange term in (\ref{BLCouloPM}) is rewritten as a 
direct product of SU$_{\spin}$(2) and SU$_{\ppin}$(2) operators,
%%%%%%%%%%%%%%%%%%%%%%%%%%%%%%%%%%%%%%%%%%%%
\begin{align}
H_{X}
= -8\sum_{\langle i,j\rangle }J_{ij}\biggl(\boldsymbol{S}(i)\!\cdot \!
\boldsymbol{S}(j)+\frac{1}{4}n(i)n(j)\biggr)
\nonumber\\  
\otimes \biggl(\sum_{a}^{x,y,z}\frac{J_{ij}^{a}}{J_{ij}}P_{a}(i)P_{a}(j)+
\frac{1}{4}n(i)n(j)\biggr)\nonumber\\
=-2\sum_{\langle i,j\rangle }\biggl(J_{ij}\boldsymbol{S}(i)\!\cdot \!
\boldsymbol{S}(j) +J_{ij}^{a}P_{a}(i)P_{a}(j)
\nonumber\\  
+J_{ij}^{b}U_{ab}(i)U_{ab}(j)+\frac{1}{4}J_{ij}n(i)n(j)\biggr),
\label{Ex}
\end{align}%
%%%%%%%%%%%%%%%%%%%%%%%%%%%%%%%%%%%%%%%%%%%%%%%
where $J_{ij}^{x}=J_{ij}^{y}=J_{ij}^{d}\equiv J_{ij}^{+}-J_{ij}^{-}$ and $%
J_{ij}^{z}=J_{ij}\equiv J_{ij}^{+}+J_{ij}^{-}$. 
The exchange integral $J_{ij}^{\pm }$ is given by 
%%%%%%%%%%%%%%%%%%%%%%%%%%%%%%%%%%%%%%%%%%%%%%%
\begin{equation}
J_{ij}^{\pm }={\frac{1}{2}}\int \!d^{2}xd^{2}y\;\varphi _{i}^{\ast }(\boldsymbol{%
x})\varphi _{j}^{\ast }(\boldsymbol{y})V_{\pm }(\boldsymbol{x}-\boldsymbol{y})\varphi
_{i}(\boldsymbol{y})\varphi _{j}(\boldsymbol{x}).
\end{equation}%
%%%%%%%%%%%%%%%%%%%%%%%%%%%%%%%%%%%%%%%%%%%%%%%%%%
In the Hamiltonian (\ref{Ex}), $\sum_{\langle i,j\rangle }$ stands
for the sum over all pairs of sites indexed by $i$ and $j$. 
 $n(i)$ is the electron number operator, 
$n\equiv \sum_{\sigma
}c_{\sigma }^{\dagger }c_{\sigma }$, at each site $i$. 
The group SU(4) is generated by the Hermitian, traceless, 
$4\times 4$ matrices, which consist of  $(4^{2}-1)$ independent generators;
 $S_{a}$, $P_{a}$, $U_{ab}\!\!\equiv\!\!2S_{a}\!\!\otimes\!\!P_{b}$.
The subgroup SU$_\text{spin}$(2)$\otimes$SU$_\text{ppin}$(2)  
is generated by $\bold{S},\bold{P}$. 
The  $U_{ab}$-generators are peculiar ones, which 
do not exist in the SU$_\text{spin}$(2)$\otimes$SU$_\text{ppin}$(2) group.  
They mix the SU$_{\text{spin}}$(2) multiplets with the 
 SU$_{\text{ppin}}$(2) ones.

The explicit forms of the SU(4) generators are given as follows.
%%%%%%%%%%%%%%%%%%%%%%%%%%%%%%%%%%%%%%%%%%%
\begin{equation}
\boldsymbol{S}=\boldsymbol{S}^{\text{f}}+\boldsymbol{S}^{\text{b}},
~~\boldsymbol{P}=\boldsymbol{P}^{\uparrow }+\boldsymbol{P}^{\downarrow }, 
\label{spinppinopera}
\end{equation}
%%%%%%%%%%%%%%%%%%%%%%%%%%%%%%%%%%%%%%%%%%%%%
where
%%%%%%%%%%%%%%%%%%%%%%%%%%%%%%%%%%%%%%%%%%%%%%%%%
\begin{align}
S_{a}^{\text{f}}& =(c_{\text{f}\uparrow }^{\dagger },c_{\text{f}\downarrow
}^{\dagger }){\frac{\tau _{a}}{2}}%
\begin{pmatrix}
c_{\text{f}\uparrow } \\ 
c_{\text{f}\downarrow }%
\end{pmatrix}%
, & \;S_{a}^{\text{b}}& =(c_{\text{b}\uparrow }^{\dagger },c_{\text{b}%
\downarrow }^{\dagger }){\frac{\tau _{a}}{2}}%
\begin{pmatrix}
c_{\text{b}\uparrow } \\ 
c_{\text{b}\downarrow }%
\end{pmatrix}%
,  \notag \\
P_{a}^{\uparrow }& =(c_{\text{f}\uparrow }^{\dagger },c_{\text{b}\uparrow
}^{\dagger }){\frac{\tau _{a}}{2}}%
\begin{pmatrix}
c_{\text{f}\uparrow } \\ 
c_{\text{b}\uparrow }%
\end{pmatrix}%
, & \;P_{a}^{\downarrow }& =(c_{\text{f}\downarrow }^{\dagger },c_{\text{b}%
\downarrow }^{\dagger }){\frac{\tau _{a}}{2}}%
\begin{pmatrix}
c_{\text{f}\downarrow } \\ 
c_{\text{b}\downarrow }%
\end{pmatrix}%
,  \label{ISN}
\end{align}%
%%%%%%%%%%%%%%%%%%%%%%%%%%%%%%%%%%%%%%%%%%%%%%%%%%
and
%%%%%%%%%%%%%%%%%%%%%%%%%%%%%%%%%%%%%%%%%%%%%%%%%%
\begin{equation}
U_{ab}= 2 S_{a}\otimes P_{b}
=\frac{1}{2}
(c_{\text{f}\uparrow }^{\dagger },c_{%
\text{f}\downarrow }^{\dagger },c_{\text{b}\uparrow }^{\dagger },c_{\text{b}%
\downarrow }^{\dagger }){{\tau _{a}}}\otimes {{\tau _{b}}}%
\begin{pmatrix}
c_{\text{f}\uparrow } \\ 
c_{\text{f}\downarrow } \\ 
c_{\text{b}\uparrow } \\ 
c_{\text{b}\downarrow }%
\end{pmatrix}.
\label{Uope}
\end{equation}%
%%%%%%%%%%%%%%%%%%%%%%%%%%%%%%%%%%%%%%%%%%%%%%%%%%%%
$c_{\sigma}(i)$ is the annihilation operator of the electron
with isospin $\sigma$ at Landau site $i$. 

In the limit $d\rightarrow 0$, which we call the SU(4)
 limit, the exchange Hamiltonian (\ref{Ex}) 
is reduced to an SU(4) invariant form,
%%%%%%%%%%%%%%%%%%%%%%%%%%%%%%%%%%%%%%%%%%%%%%
\begin{align}
&H_{X}\rightarrow H_{X}^{+}
=- 2\sum_{\langle i,j\rangle }J_{ij}^{+}\biggl(\boldsymbol{S}
(i)\!\cdot \!\boldsymbol{S}(j)+\boldsymbol{P}(i)\!\cdot \!\boldsymbol{P}(j)  
\nonumber\\
&~~~~~~~~~~~~~~~~~~~~~~~ +U_{ab}(i)\!\cdot \!U_{ab}(j)
+{\frac{1}{4}}n(i)n(j)\biggr).
\label{Exx}
\end{align}%
%%%%%%%%%%%%%%%%%%%%%%%%%%%%%%%%%%%%%%%%%%%%%%%%
In the SU(4) limit, the interlayer Coulomb potential V$_\text{fb}$ 
(V$_\text{bf}$) goes to 
the intralayer Coulomb potential V$_\text{ff}$ (V$_\text{bb}$).
Hence SU(4)-noninvariant Coulomb interaction (\ref{BBLCouloM}) vanishes. 
The SU(4)-invariant exchange Hamiltonian (\ref{Exx}) is obtained 
  from $H_{C}^+$.
 
Meanwhile in the limit $d\rightarrow \infty$ where two layers are sufficiently 
separated, 
the pseudospin stiffness $J_d$ goes to zero.
 The exchange Hamiltonian (\ref{Ex}) is reduced to the Hamiltonian used in
 Demler and Das Sarma \cite{c4}.
%%%%%%%%%%%%%%%%%%%%%%%%%%%%%%%%%%%%%%%%%%%%%%
\begin{align}
&H_{X}\rightarrow -8\sum_{\langle i,j\rangle } J_{ij}\bigl(\boldsymbol{S}(i)\!\cdot
\!\boldsymbol{S}(j)+{\frac{1}{4}}n(i)n(j)\bigr)  
\nonumber\\
&~~~~~~~~~~~~~~~~~~~~~~\otimes \bigl({P_z}(i)\!\cdot \!{P_z}(j)+{\frac{1}{4}}n(i)n(j)
\bigr)
\nonumber\\
&~~~~~~~=- 4\sum_{\langle i,j\rangle }J_{ij}\biggl
(\boldsymbol{S}^{\text{f}}(i)\!\cdot \!\boldsymbol{S}^{\text{f}}(j)+\boldsymbol{S}^{\text{b}}(i)\!\cdot \!\boldsymbol{S}^{\text{b}}(j)  
\nonumber\\
&~~~~~~~~
~~~~~~~~~~~~+\frac{1}{4}n_f(i)n_f(j)+\frac{1}{4}n_b(i)n_b(j)\biggr),
\end{align}%
%%%%%%%%%%%%%%%%%%%%%%%%%%%%%%%%%%%%%%%%%%%%%%%%
where the symmetry of the exchange interaction  is 
SU$_\text{front}$(2)$\otimes$SU$_\text{back}$(2).
Each SU(2) represents  the spin rotation symmetry on each  layer.

The direct interaction is given  as 
%%%%%%%%%%%%%%%%%%%%%%%%%%%%%%%%%%%%%%%%%%%%%%%%%%
\begin{equation}
H_{D}=\sum_{i=1}^{N_{\Phi }}\biggl(-{{\Delta }_{\text{Z}}}S_{z}(i)+{{%
\varepsilon }_{\text{cap}}}P_{z}(i)P_{z}(i)-{{\Delta }_{\text{SAS}}}%
P_{x}(i)\biggr),  \label{BDct}
\end{equation}%
%%%%%%%%%%%%%%%%%%%%%%%%%%%%%%%%%%%%%%%%%%%%%%%%%%%%%%
where ${\Delta }_{\text{Z}}$ and ${\Delta }_{\text{SAS}}$ are
the Zeeman and tunneling gaps, while $\varepsilon_{\text{cap}}$
 is the capacitance energy,
%%%%%%%%%%%%%%%%%%%%%%%%%%%%%%%%%%%%%%%%%%%%%%%
\begin{equation}
\varepsilon _{\text{cap}}=\frac{e^{2}}{4\pi \varepsilon \ell _{B}}\sqrt{%
\frac{\pi }{2}}\biggl(1-e^{d^{2}/2\ell _{B}^{2}}\bigl\{1-\text{erf}\bigl(d/%
\sqrt{2}\ell _{B})\bigr\}\biggr).  \label{veps}
\end{equation}%
%%%%%%%%%%%%%%%%%%%%%%%%%%%%%%%%%%%%%%%%%%%%%%%

The total Landau-site Hamiltonian is 
 the sum of the direct term (\ref{BDct}) and the exchange term (\ref{Ex}),
%%%%%%%%%%%%%%%%%%%%%%%%%%%%%%%%%%%%%%%%%%%%
\begin{equation}
H_{\text{tot}}=H_{D}+H_{X}.  \label{TotalHamil}
\end{equation}%
%%%%%%%%%%%%%%%%%%%%%%%%%%%%%%%%%%%%%%%%%%%%

\section{Ground States}\label{SecGrounState}

At $\nu =2$,  one Landau site contains two
electrons, each of which belongs to the $\boldsymbol{4}$-dimensional irreducible
representation of SU(4). 
They  form a Schwinger boson \cite{SB}, which belongs
 to the $\boldsymbol{6}$-dimensional irreducible representation due to 
their antisymmetricity,
\begin{equation}
(\boldsymbol{4}\otimes \boldsymbol{4})_{\text{antisymmetric}}=
\boldsymbol{6}.
\label{44to106}
\end{equation}%
In the language of the subgroup SU$_{\text{spin}}$(2)$\otimes $SU$_{\text{%
ppin}}$(2), the $\boldsymbol{6}$-dimensional irreducible
representation 
is divided into two different irreducible representations
%%%%%%%%%%%%%%%%%%%%%%%%%%%%%%%%%%%%%%%%%%%%%%
\begin{equation}
\boldsymbol{6}
=(\boldsymbol{3},\boldsymbol{1})+(\boldsymbol{1},\boldsymbol{3}),
\end{equation}%
%%%%%%%%%%%%%%%%%%%%%%%%%%%%%%%%%%%%%%%%%%%%
where $\boldsymbol{3}$ is the symmetric representation of SU(2), and $\boldsymbol{1}$
is the antisymmetric representation of SU(2).
The $(\boldsymbol{3},\boldsymbol{1})$ sector is called 
a spin-sector and the $(\boldsymbol{1},\boldsymbol{3})$ sector a 
ppin-sector.
The SU$_\text{spin}$(2) and SU$_\text{ppin}$(2) operators 
(\ref{spinppinopera}) rotate  the 
spin-  and ppin-sectors individually, however they do not mix both sectors.
Only the $U$-operators can mix them.  

\subsection{Spin phase and Ppin phase}
The ground state and its energy in each sector are obtained by
minimizing the  Hamiltonian (\ref{TotalHamil}). 

The spin-sector $(\boldsymbol{3},\boldsymbol{1})$ consists of spin-triplet
pseudospin-singlet states, 
%%%%%%%%%%%%%%%%%%%%%%%%%%%%%%%%%%%%%%%%%%%%%%%
\begin{align}
& |t_{\uparrow }\rangle =|\text{f}^{\uparrow },\text{b}^{\uparrow }\rangle
,\quad \quad |t_{0}\rangle ={\frac{1}{\sqrt{2}}}\bigl(|\text{f}^{\uparrow },%
\text{b}^{\downarrow }\rangle +|\text{f}^{\downarrow },\text{b}^{\uparrow
}\rangle \bigr),  \notag \\
& |t_{\downarrow }\rangle =|\text{f}^{\downarrow },\text{b}^{\downarrow
}\rangle .  \label{GrounT}
\end{align}%
%%%%%%%%%%%%%%%%%%%%%%%%%%%%%%%%%%%%%%%%%%%%%%%%%%%
$t_{a}(i)$'s satisfy hardcore bosonic commutation relations \cite{SB}. We call
them $t$-bosons.

The ppin-sector $(\boldsymbol{1},\boldsymbol{3})$ consists of spin-singlet
pseudospin-triplet states, 
%%%%%%%%%%%%%%%%%%%%%%%%%%%%%%%%%%%%%%%%%%%%%%%%%
\begin{align}
&|\tau _{+}\rangle =|\text{f}^{\uparrow },\text{f}^{\downarrow }\rangle
,\quad \quad |\tau _{0}\rangle =\frac{1}{\sqrt{2}}\bigl(|\text{f}^{\uparrow
},\text{b}^{\downarrow }\rangle -|\text{f}^{\downarrow },\text{b}^{\uparrow
}\rangle \bigr),  \notag \\
& |\tau _{-}\rangle =|\text{b}^{\uparrow },\text{b}^{\downarrow }\rangle .
\label{GrounTau}
\end{align}
%%%%%%%%%%%%%%%%%%%%%%%%%%%%%%%%%%%%%%%%%%%%%%%%%%%%%%%%
$\tau _{a}(i)$'s also satisfy hardcore 
bosonic commutation relations. We
call them $\tau $-bosons.

In the spin-sector, the minimum energy is  given by 
%%%%%%%%%%%%%%%%%%%%%%%%%%%%%%%%%%%%%%%%%%%%%%%
\begin{equation}
E_{t_{\uparrow }} =-\Delta _{Z}-2J,  \label{Spinphene} 
\end{equation}
%%%%%%%%%%%%%%%%%%%%%%%%%%%%%%%%%%%%%%%%%%%%%%%%%%
where the spin stiffness $J$ is given as 
%%%%%%%%%%%%%%%%%%%%%%%%%%%
\begin{equation}
J = \frac{1} {16\sqrt{2\pi}}\frac{e^{2}}{4\pi\varepsilon \ell _{B}}.
\label{IntroSpinStiff}
\end{equation}
%%%%%%%%%%%%%%%%%%%%%%%%%%%
The ground state of the spin-sector is $\prod_{i=1}^{N_{\Phi }}|t_{\uparrow
}\rangle _{i}$ with the ground-state energy $N_{\Phi }E_{t_{\uparrow }}$.

In the ppin-sector, the minimum energy state is
%%%%%%%%%%%%%%%%%%%%%%%%%%%%%%%%%%%%%%%%%%%%%%%%%%%%%%%%%%%
\begin{equation}
|v_{+}\rangle  =\frac{\cos \theta -\sin \theta }{2}(|\tau _{+}\rangle
+|\tau _{-}\rangle )+\frac{\cos \theta +\sin \theta }{\sqrt{2}}|\tau
_{0}\rangle ,   \label{TrueGrounP} 
\end{equation}
%%%%%%%%%%%%%%%%%%%%%%%%%%%%%%%%%%%%%%%%%%%%%%%%%%%
where 
%%%%%%%%%%%%%%%%%%%%%%%%%%%%%%%%%%%%%%%%%%%%%%%%%%%
\begin{equation}
\tan \theta =\frac{\varepsilon _{\text{cap}}}{2{\Delta }_{\text{SAS}}+%
\sqrt{4{\Delta }_{\text{SAS}}^{2}+\varepsilon _{\text{cap}}^{2}}},
\end{equation}%
%%%%%%%%%%%%%%%%%%%%%%%%%%%%%%%%%%%%%%%%%%%%%%%%%%%%%%%%
with the energy 
%%%%%%%%%%%%%%%%%%%%%%%%%%%%%%%%%%%%%%%%%%%%%%%%%%%%%%%%
\begin{equation}
E_{v_{+}} ={\frac{1}{2}}\bigl(\varepsilon _{%
\text{cap}}-\sqrt{4{\Delta }_{\text{SAS}}^{2}+\varepsilon _{\text{cap}%
}^{2}}\bigr)-(J+J^{d}\cos ^{2}(2\theta )). \label{TrueEnergP}
\end{equation}
%%%%%%%%%%%%%%%%%%%%%%%%%%%%%%%%%%%%%%%%%%%%%%%%%%%%%%
The  pseudospin stiffness $J^d$ is given as 
%%%%%%%%%%%%%%%%%%%%%%%%%%%
\begin{equation}
{\frac{J^{d}}{J}} = -\sqrt {\frac{2}{\pi}}{\frac{d}{\ell _{B}}}
 +\biggl(1+{\frac{d^{2}}{\ell _{B}^{2}}}\biggr)e^{d^2/2\ell _B^2}\biggl(1-\text{erf}\bigl(d/\sqrt {2}\ell _{B}\bigr)\biggr).
\end{equation}
%%%%%%%%%%%%%%%%%%%%%%%%%%%
 The ground state of the ppin-sector is $\prod_{i=1}^{N_{\Phi
}}|v_{+}\rangle _{i}$ with the ground-state energy $N_{\Phi }E_{v_{+}}$.
With the decrease of the interlayer separation $d$, 
 the $v_{+}$-boson exchange energy  also decreases,
 while the $t_{\upA}$-boson exchange energy does not change keeping the
minimum value $-2J$. 
The $v_{+}$-boson exchange energy
 yields the minimum  value $-2J$ only when  two layers coincide.

Consequently, there are two possible ground states,
 $|\Phi_{S}\rangle=\prod_{i=1}^{N_{\Phi}}|t_{\uparrow}\rangle _{i}$
 or $|\Phi_{P}\rangle=\prod_{i=1}^{N_{\Phi}}|v_{+}\rangle_{i}$.
When $E_{t_{\uparrow }}<E_{v_{+}}$, the ground state is $|\Phi_{S}\rangle$,
 which we call the spin phase since all
spins are polarized. On the other hand, when $E_{v_{+}}<E_{t_{\uparrow}}$,
the ground state is $|\Phi_{P}\rangle$, which we
call the ppin phase.

\subsection{C-phase}\label{C-phaseene}
 
Instead of  the SU(4) operators given in (\ref{spinppinopera}) and 
(\ref{Uope}),
 it is convenient to use such 
  raising, lowering operators,
%%%%%%%%%%%%%%%%%%%%%%%%%%%%%%%%%%%%%%%%%%%%%%%
\begin{align}
&S_+ = \frac{1}{\sqrt{2}}(S_x+iS_y),~~S_- = \frac{1}{\sqrt{2}}(S_x-iS_y),
\nonumber\\
&P_+ = \frac{1}{\sqrt{2}}(P_x+iP_y),~~P_- = \frac{1}{\sqrt{2}}(P_x-iP_y),
\nonumber\\
&U_{++}= 2 S_+ P_+,~~~~~~~~~~U_{+-}= 2 S_+ P_-,\nonumber\\ 
&U_{-+}= 2 S_- P_+,~~~~~~~~~~U_{--}= 2 S_- P_-,\nonumber\\
&U_{z+}= 2 S_z P_+,~~~~~~~~~~~U_{+z}=2  S_+ P_z,\nonumber\\
&U_{z-}= 2 S_z P_-,~~~~~~~~~~~U_{-z}=2 S_- P_z,\nonumber\\
\end{align}
%%%%%%%%%%%%%%%%%%%%%%%%%%%%%%%%%%%%%%%%%%%%%%%%%%%
 and  Cartan operators; $S_z$, $P_z$, $U_{zz}\!=\!2 S_z\otimes P_z$.
  $S_{+(-)}$ raises (lowers) the spin $z$-component by one; 
$ S_{+(-)}|t_0\rangle = |t_{\upA(\dnA)}\rangle$.
Similarly, $P_{+(-)}$ raises (lowers) the ppin $z$-component by 
one;
$ P_{+(-)}|\tau_0\rangle = |\tau_{+(-)}\rangle$.
 $U_{++},U_{+-},U_{-+}$ and $U_{--} $ change not only
 the  spin but also the ppin  $z$-component 
 by one and mix 
 the spin-sector with the  ppin-sector, for instance
$ U_{-+}|t_{\upA}\rangle = |\tau_+\rangle$.
The $U_{++},U_{+-}$ and $U_{+z}$  change
the ppin-sector states
to the spin-sector states,
 while the  $U_{--},U_{-+}$ and $U_{-z}$  do the 
spin-sector states to the ppin-sector states.

 We focus on a Hilbert space  
 spanned by the lowest energy states $|t_{\upA}\rangle,|v_+\rangle$ 
 and search for the lowest energy in this subspace.
 We abbreviate $|t_{\upA}\rangle,|v_+\rangle$ as 
 $|t\rangle,|v\rangle$ for simplicity. 
 The effective  one body 
 operator which operates on this subspace is represented as  
%%%%%%%%%%%%%%%%%%%%%%%%%%%
\begin{equation}
 O^{eff} = \sum _{i}\langle b_i|O(i)|b'_{i}\rangle b^{\dagger}_i  b'_i,
\end{equation}
%%%%%%%%%%%%%%%%%%%%%%%%%%%
where  $b\!\!=\!\!t,v$.
The   matrix representations of the SU(4) operators  $S_a$,$P_a$,$U_{ab}$ 
 are   
%%%%%%%%%%%%%%%%%%%%%%%%%%%%
\begin{subequations}
\begin{align}
&S_z
\circeq\begin{pmatrix}
   1    &     0  \\
     0 &     0
\end{pmatrix}
,\\
&P_+ \circeq P^{\dagger}_-
\circeq\frac{\cos(2\theta)}{\sqrt{2}}
\begin{pmatrix}
     0  &     0  \\
     0 &     1
\end{pmatrix} 
,~~~~~~~~~~~~~~~~\\
&U_{++}\circeq U^{\dagger}_{--}\circeq -U_{+-}\circeq -U^{\dagger}_{-+}
\circeq\frac{\cos\theta-\sin\theta}{2}
\begin{pmatrix}
   0   &     -1  \\
     0 &     0
\end{pmatrix} 
,~~~~~~~~~~~~~~~\label{U++etc}\\
&U_{+z} \circeq U^{\dagger}_{-z}
\circeq\frac{\cos\theta +\sin\theta }{\sqrt{2}}
\begin{pmatrix}
     0  &     -1  \\
     0 &     0
\end{pmatrix}.\label{U+zetc}\nonumber\\
\end{align}
\label{matrixSU(4)}
\end{subequations}
%%%%%%%%%%%%%%%%%%%%% 
Matrices of other SU(4) operators are zeros. 
Due to the mixing of the spin- and the ppin-sectors
 by $U$-operators,  only the $U_{ab}$ matrices  
  give off-diagonal elements,
 which  represent the interchange of $t$ and $v$-bosons.
The matrix of the  direct term (\ref{BDct}) is given as 
%%%%%%%%%%%%%%%%%%%%%%%%%%%%
\begin{equation}
H_D
\circeq
\begin{pmatrix}
 E^{D}_t &\!\!0     \\
 \!\! 0     &\!\!    E^D_v  \\
\end{pmatrix},
\end{equation}
%%%%%%%%%%%%%%%%%%%%% 
 where
%%%%%%%%%%%%%%%%%%%%%%%%%%%%%%%%%%%%%%%  
\begin{subequations}
\begin{align}
&E_t^D=-\Delta_{\text{Z}},\\
&E_v^D=-\sqrt{\Delta_{\text{SAS}}^2
+(\frac{\varepsilon_{\text{cap}}}{2})^2}+\frac{\varepsilon_{\text{cap}}}{2}.
\end{align} 
\end{subequations}
%%%%%%%%%%%%%%%%%%%%%%%%%%%%%%%%%%%%%%%%
Next, we consider the matrix elements for the 
 exchange interaction term (\ref{Ex}).
 First we rewrite the exchange interaction term with the use of raising and
 lowering operators.
The spin-spin interaction  is rewritten as 
%%%%%%%%%%%%%%%%%%%%%%%%%%%%%%%%%%%%%%%%%%%%%%%
\begin{align}
&\boldsymbol{S}(i)\cdot \boldsymbol{S}(j)
\nonumber\\&=S_+(i)S_-(j)+ S_-(i)S_+(j) +S_z(i)S_z(j).
\end{align}
%%%%%%%%%%%%%%%%%%%%%%%%%%%%%%%%%%%%%%%%%%%%%%%%%%%
Similarly the ppin-ppin  interaction  is
%%%%%%%%%%%%%%%%%%%%%%%%%%%%%%%%%%%%%%%%%%%%%%%
\begin{align}
&\boldsymbol{P}(i)\cdot \boldsymbol{P}(j)
\nonumber\\&=P_+(i)P_-(j)+ P_-(i)P_+(j) +P_z(i)P_z(j).
\end{align}
%%%%%%%%%%%%%%%%%%%%%%%%%%%%%%%%%%%%%%%%%%%%%%%%%%%
The other exchange interaction  is  
%%%%%%%%%%%%%%%%%%%%%%%%%%%%%%%%%%%%%%%%%%%%%%%
\begin{align}
& {U}_{ab}(i)\cdot{U}_{ab}(j)
\nonumber\\&=  U_{++}(i)U_{--}(j)+ U_{+-}(i)U_{-+}(j)
\nonumber\\&+ U_{-+}(i)U_{+-}(j)+ U_{--}(i)U_{++}(j)
\nonumber\\&+ U_{z+}(i)U_{z-}(j)~+ U_{z-}(i)U_{z+}(j)
\nonumber\\&+ U_{+z}(i)U_{-z}(j)~+ U_{-z}(i)U_{+z}(j)
\nonumber\\&+ U_{zz}(i)U_{zz}(j).
\end{align}
%%%%%%%%%%%%%%%%%%%%%%%%%%%%%%%%%%%%%%%%%%%%%%%%%%%
To reproduce the exchange interaction (\ref{Ex}), the spin and 
the ppin stiffness are needed in front of these operators.
Each of   $ U_{++},U_{--},U_{+-},U_{-+}$ exchange interaction parts  needs
  the ppin stiffness $J^d$ as its coefficient,
 while  each of $U_{z+},U_{z-},U_{+z},U_{-z},U_{zz}$ parts does the 
  spin stiffness $J$.
 The difference between these two coefficients 
 represents the magnitude of the  SU(4)-noninvariant interaction.

 Two body operators are effectively described as 
%%%%%%%%%%%%%%%%%%%%%%%%%%%
\begin{align}
&O^{eff} = \sum _{ij}\langle  b_j  b_i|O (i) \tilde{O} (j)
|b'_{i} b'_{j}\rangle~b^{\dagger}_j b'_j  b^{\dagger }_i b'_i \nonumber\\
&~~~~~~~ = \sum _{ij}\langle   b_i|O (i)|b'_{i} \rangle
\otimes\langle   b_j|\tilde{O} (j)|b'_{j} \rangle
 ~b^{\dagger}_j b'_j  b^{\dagger }_i b'_i.
\end{align}
%%%%%%%%%%%%%%%%%%%%%%%%%%%
%where the bosonic operators $b,b'$ are  $b,b'=t,v$.
One can easily obtain  matrix elements of 
two body operators from
the direct product 
 of  one body operators.
Using the basis of two body  Hilbert space; $|tt\rangle,|tv\rangle,|vt\rangle,
 |vv\rangle$, we take  matrix elements 
of the exchange interaction term (\ref{Ex}).
With the use of (\ref{matrixSU(4)}), the matrix  
of the exchange interaction  is written as,
%%%%%%%%%%%%%%%%%%%%%%%%%%%%
\begin{align}
&H_X
\circeq -J
\nonumber\\
&-\begin{pmatrix}
 J &\!\!0   &\!\! 0  &\!\!0  \\
 \!\! 0     &\!\!    0  &J^{+}+J^{-}\sin(2\theta) &\!\!0  \\
 \!\!    0  &J^{+}+J^{-}\sin(2\theta)   &\!\! 0 &\!\! 0 \\
 \!\!    0  &\!\!0  &\!\! 0 & J^{d}\cos^2(2\theta)
\end{pmatrix} ,
\label{Exmatrix}
\end{align}
%%%%%%%%%%%%%%%%%%%%% 
with $J^{\pm}\equiv\frac{1}{2}(J\pm J^d)$.
The first term  $-J$ comes from the density-density interaction term. 
The (1,1) and  (4,4) diagonal components in the matrix come  from 
the spin and the ppin
 exchange interactions individually.
Meanwhile the off-diagonal components come from the  SU(4) peculiar 
 exchange term $U_{ab}(i)U_{ab}(j)$.
These off-diagonal elements yield an origin of the C-phase.
  
\begin{figure}[htbp]
  \begin{center}
   \scalebox{0.45}{\includegraphics{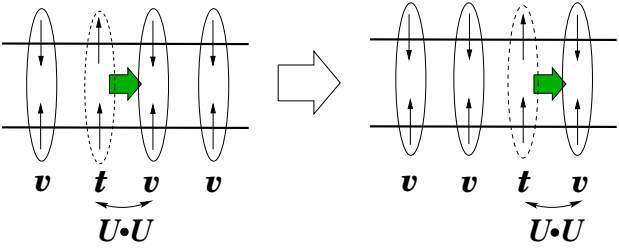}}
  \end{center}
  \caption {A $t$-boson hops to the
 nearest neighbor site by the SU(4) peculiar exchange interactions stemmed 
 from $U$-operators.}  
   \label{Coherent}
\end{figure}

Consequently, we obtain the effective Hamiltonian, 
%%%%%%%%%%%%%%%%%%%%%%%%%%%%
\begin{align}
&H^{eff}_{\text{tot}}=
 E_t^D\sum_i t^{\dagger}_i t_i + E_v^D\sum_i v^{\dagger}_i v_i-J
\nonumber\\
&~~~~~~~~~~- J\sum_{<ij>} t^{\dagger}_j t_j t^{\dagger}_i t_i 
-J^{d}\cos^2(2\theta)\sum_{<ij>} v^{\dagger}_j  v_j v^{\dagger}_i v_i
\nonumber\\
&~~~~~~~~~~-(J^{+}+J^{-}\sin(2\theta))
 \sum_{<ij>}(v^{\dagger}_j t_j  t^{\dagger}_i v_i 
   +t^{\dagger}_j v_j v^{\dagger}_i    t_i).
\label{nrh}
\end{align}
%%%%%%%%%%%%%%%%%%%%%%%%%%%% 
The last  term which 
comes from the off-diagonal elements of the $U$-operators
 flips the $t$-boson to the $v$-boson and vice versa. 
Due to this effect,  $t$-bosons are  distributed  
in the $v$-boson condensation state [Fig.\ref{Coherent}]. 
%%%%%%%%%%%%%%%%%%%%%%%%%%%%%%%%%%%%%%%%%%%%%%
%%%%%%%%%%%%%%%%%%%%%%%%%%%%%%%%%%%%%%%%%%%%%%%

Using the variation method, we find out the ground  states of 
the effective Hamiltonian.
The variational function is    
%%%%%%%%%%%%%%%%%%%%%%%%%%%% 
\begin{equation}
|\Phi\rangle =\prod_{i}[\alpha t^{\dagger}+\beta  v^{\dagger} ]_{i}
      |0\rangle .
\label{vf}
\end{equation}
%%%%%%%%%%%%%%%%%%%%%%%%%%%% 
Due to the normalization condition, parameters $\alpha,\beta$ must satisfy such  constraint,
%%%%%%%%%%%%%%%%%%%%%%%%%%%% 
\begin{equation}
 |\alpha|^2 + |\beta|^2 =1 .
\label{cc}
\end{equation}
%%%%%%%%%%%%%%%%%%%%%%%%%%%% 
The expectation value of the reduced Hamiltonian ($\ref{nrh}$) reads 
%%%%%%%%%%%%%%%%%%%%%%%%%%%% 
\begin{align}
&E =\langle \Phi | H^{eff}_{\text{tot}} |\Phi \rangle  \nonumber\\
&~~=E^D_t|\alpha|^2 + E^D_v|\beta|^2 -J
  - J|\alpha|^4\nonumber\\
&~~~~-J^{d}\cos^2(2\theta)|\beta|^4 -2(J^{+}+J^{-}\sin(2\theta))|\alpha|^2 |\beta|^2 ,
\label{en}
\end{align}
%%%%%%%%%%%%%%%%%%%%%%%%%%%% 
which coincides with 
the energy in the Demler and Das Sarma \cite{c4} in the limit 
$d\!\rightarrow\!\infty$,
%%%%%%%%%%%%%%%%%%%%%%%%%%%% 
\begin{equation}
E 
\rightarrow E^D_t|\alpha|^2 + E^D_v|\beta|^2 -J
  - J|\alpha|^4 -J(1+\sin(2\theta))|\alpha|^2 |\beta|^2.
\\
\end{equation} 
%%%%%%%%%%%%%%%%%%%%%%%%%%%% 
Meanwhile in the SU(4) limit $d\!\rightarrow\! 0$,
 it yields 
%%%%%%%%%%%%%%%%%%%%%%%%%%%% 
\begin{equation}
E \rightarrow E^D_t|\alpha|^2 + E^D_v|\beta|^2 -2J,
\end{equation}
%%%%%%%%%%%%%%%%%%%%%%%%%%%% 
 where we have used the constraint (\ref{cc}).
In this SU(4) limit, only  two ground states are allowed.
One is  the case $(|\alpha|,|\beta|)=(1,0)$, where 
the ground state energy reads
%%%%%%%%%%%%%%%%%%%%%%%%%%%% 
\begin{equation}
E = -\Delta_{\text{Z}} -2J.
\end{equation}
%%%%%%%%%%%%%%%%%%%%%%%%%%%%
 This is the spin phase. 
The other case is  $(|\alpha|,|\beta|)=(0,1)$, where 
 the ground state  energy reads 
%%%%%%%%%%%%%%%%%%%%%%%%%%%% 
\begin{equation}
E =-\Delta_{\text{SAS}} -2J.
\end{equation}
%%%%%%%%%%%%%%%%%%%%%%%%%%%% 
 This is the ppin phase.
Note the exchange energies are degenerate in two phases,
 because  the exchange term becomes 
the SU(4) Casimir operator in the SU(4) limit. 

 Under the constraint ($\ref{cc}$), we minimize the energy ($\ref{en}$). 
There arise three different kinds of  ground states,
%%%%%%%%%%%%%%%%%%%%%%%%%%%% 
\begin{subequations}
\begin{align}
&  (|\alpha|,|\beta|)=(1,0)~~~~ if ~~t_{min}<0,\label{firstph} \\ 
&(|\alpha|,|\beta|)=(\sqrt{1-t_{min}},\sqrt{t_{min}})~~ if~~ 0<t_{min} <1,
\label{secondph} \\
&(|\alpha|,|\beta|)=(0,1)~~~~ if~~ 1< t_{min}  ,\label{thirdph}
\end{align}
\end{subequations}
%%%%%%%%%%%%%%%%%%%%%%%%%%%% 
where
%%%%%%%%%%%%%%%%%%%%%%%%%%%% 
\begin{equation}
t_{min} =\frac{E_t^D -E_v^D -2J^-(1-\sin(2\theta))}
      {4J^{-}\sin(2\theta)+2J^{d}\sin^2(2\theta)} .
\label{tmin}
\end{equation}
%%%%%%%%%%%%%%%%%%%%%%%%%%%% 
With the use of  $t_{min}$, the energy  
is given as 
%%%%%%%%%%%%%%%%%%%%%%%%%%
\begin{equation}
E=E_t^D-2J-(2J^-\sin(2\theta)+J^d\sin^2(2\theta))t^2_{min}.
\label{ThreeG}
\end{equation}
%%%%%%%%%%%%%%%%%%%%%%%%%%%%%

At $t_{min}=0$, the energy (\ref{ThreeG})
 coincides with the spin phase energy (\ref{Spinphene}).
The wave function reads 
%%%%%%%%%%%%%%%%%%%%%%%%%%%% 
\begin{equation} 
   |\Phi_{S}\rangle =\prod_{i}t^{\dagger}_i|0\rangle .
\end{equation}
%%%%%%%%%%%%%%%%%%%%%%%%%%%% 
Therefore,  the spin phase can be interpreted as a $t$-boson condensation phase.

At $t_{min}=1$, the energy (\ref{ThreeG})
 coincides with the ppin phase energy (\ref{TrueEnergP}).
The wave function reads 
%%%%%%%%%%%%%%%%%%%%%%%%%%%% 
\begin{equation} 
   |\Phi_{P}\rangle =\prod_{i}v^{\dagger}_i|0\rangle .
\end{equation}
%%%%%%%%%%%%%%%%%%%%%%%%%%%% 
The ppin phase is   a  $v$-boson condensation phase.

The last case (\ref{secondph}) is the C-phase, where 
the wave function  is given as  
%%%%%%%%%%%%%%%%%%%%%%%%%%%% 
\begin{equation} 
 |\Phi_C \rangle =\prod_{i}[\alpha t^{\dagger}
     +\beta v^{\dagger} ]_i|0\rangle ,
\label{Cphasewave}
\end{equation}
%%%%%%%%%%%%%%%%%%%%%%%%%%%% 
with the energy (\ref{ThreeG}).
Both $t$-bosons and $v$-bosons  
simultaneously condense in this phase.

 To understand   magnetic properties in each phase,
 we study a spin expectation value in each layer.
The $t$-bosons  are the highest weight 
 states in the spin triplet.
Therefore, in the spin phase, the spins in both layers  are fully polarized,
%%%%%%%%%%%%%%%%%%%%%%%%%%%% 
\begin{subequations}
\begin{align}
&\langle\Phi_{S}|\boldsymbol{S}^{\text{f}}|\Phi_{S}\rangle =\frac{1}{2}(0,0,1)
 \label{SSfS} ,\\
&\langle\Phi_{S}|\boldsymbol{S}^{\text{b}}|\Phi_{S}\rangle =\frac{1}{2}(0,0,1)
 \label{SSbS} .
\end{align}
\end{subequations}
%%%%%%%%%%%%%%%%%%%%%%%%%%%% 
The $v$-bosons  condense, which are  spin singlet states.
 Hence, in the ppin phase,  no  spins are    polarized at all, where
 the expectation values of the spin operators in the front layer
  and in the back layer  are individually 0,    
%%%%%%%%%%%%%%%%%%%%%%%%%%%% 
\begin{subequations} 
\begin{align}
&\langle\Phi_{P}|\boldsymbol{S}^{\text{f}}|\Phi_{P}\rangle =(0,0,0) \label{PSfP} ,\\
&\langle\Phi_{P}|\boldsymbol{S}^{\text{b}}|\Phi_{P}\rangle =(0,0,0) \label{PSbP} .
\end{align}
\end{subequations}
%%%%%%%%%%%%%%%%%%%%%%%%%%%% 
In the C-phase, the $t$-bosons and  $v$-bosons simultaneously condense.
With the use of Eq.(\ref{matrixSU(4)}), 
expectation values of SU(4) isospins with the C-phase wave 
function (\ref{Cphasewave}) are  given as 
%%%%%%%%%%%%%%%%%%%%%%%%%%%% 
\begin{align}
&\langle{S}_z\rangle_C=|\alpha|^2, ~~~~~~~
\langle {P}_x \rangle_C 
 = \cos(2\theta)|\beta|^2, \nonumber\\
&\langle{U_{xy}}\rangle_C
  =-(\cos\theta-\sin\theta)\text{Im}(\alpha^* \beta),\nonumber\\
&\langle{U_{xz}}\rangle_C
  =-(\cos\theta+\sin\theta)\text{Re}(\alpha^* \beta),\nonumber\\
&\langle{U_{yy}}\rangle_C
  =(\cos\theta-\sin\theta)\text{Re}(\alpha^* \beta),\nonumber\\
&\langle{U_{yz}}\rangle_C
  =-(\cos\theta+\sin\theta)\text{Im}(\alpha^* \beta).\nonumber\\
\label{SU4C}
\end{align}
%%%%%%%%%%%%%%%%%%%%%%%%%%%% 
%%%%%%%%%%%%%%%%%%%%%%%%%%%%%%%%%%%%%%%%%
Other expectation values of SU(4) isospins are zeros.
%From the equations (\ref{SU4C}), 
$U$-operators have non-zero values only in the C-phase,
which become a N\'{e}el order parameter as we shall see below [TABLE \ref{order}].

The spin operators in two layers are given as  
%%%%%%%%%%%%%%%%%%%%%%%%%%%%%%%%%%%%%%%%%%%%
\begin{subequations}
\begin{align}
&\langle\boldsymbol{S}^{{\text{f}}}\rangle_C 
=\frac{1}{2}(0,0,|\alpha|^2)-\frac{\cos\theta+\sin\theta}{2}
              (\text{Re}(\alpha^*\beta),
               \text{Im}(\alpha^*\beta),0),              
               \label{CSfC}\\
&\langle\boldsymbol{S}^{{\text{b}}}\rangle_C 
=\frac{1}{2}(0,0,|\alpha|^2)+\frac{\cos\theta+\sin\theta}{2}
              (\text{Re}(\alpha^*\beta),
               \text{Im}(\alpha^*\beta),0) ,             
               \label{CSbC}
\end{align}
\end{subequations}
%%%%%%%%%%%%%%%%%%%%%%%%%%%%%%%%%%%%%%%%%%%%%%%
where we have used the relations, $S_a^{\text{f}}=\frac{1}{2}(S_a+U_{az}), S_a^{\text{b}}=\frac{1}{2}(S_a-U_{az})$.
 The C-phase exhibits   antiferromagnetic  properties
 in the two dimensional layer directions, while  ferromagnetic properties  
 in the perpendicular direction.

The spin N\'{e}el  parameter 
$ \boldsymbol{N}_{spin}= \boldsymbol{S}^{\text{f}} - \boldsymbol{S}^{\text{b}}$,
 which is equivalent to $U$-operator $U_{az}$,
is introduced as an order parameter of the C-phase.
In the spin and ppin phases, its expectation values  are trivial; 
 $\langle\Phi_{S}|\boldsymbol{N}_{spin}|\Phi_{S}\rangle =
\langle\Phi_{P}|\boldsymbol{N}_{spin}|\Phi_{P}\rangle =\bold{0}$,
 while in the C-phase, it yields a nonzero value,
%%%%%%%%%%%%%%%%%%%%%%%%%%%%%%%%%%%%%%%%%%%%%%
\begin{equation}
\langle\boldsymbol{N}_{spin}\rangle_C
 =-(\cos\theta+\sin\theta)
              (\text{Re}(\alpha^*\beta),
             \text{Im}(\alpha^*\beta),0).\label{CNC}
\end{equation}
%%%%%%%%%%%%%%%%%%%%%%%%%%%%%%%%%%%%%%%%%%%%%%%%%%%%
Thus, $ \boldsymbol{N}_{spin}$ takes a finite value only in the C-phase, which 
 represents an antiferromagnetic property of spins.

The ppin operators on  $x$-polarized two
 spin-states $\upA_x, \dnA_x$ are given as  
%%%%%%%%%%%%%%%%%%%%%%%%%%%%%%%%%%%%%%%%%%%%
\begin{subequations}
\begin{align}
&\langle\boldsymbol{P}^{{\upA_x}}\rangle_C =
\frac{1}{2}\cos(2\theta)(|\beta|^2,0,0)\nonumber\\
&-\frac{1}{2}
              (0, (\cos\theta-\sin\theta)    \text{Im}(\alpha^*\beta),
              (\cos\theta+\sin\theta) \text{Re}(\alpha^*\beta)),
\\              
&\langle\boldsymbol{P}^{{\dnA_x}}\rangle_C =
\frac{1}{2}\cos(2\theta)(|\beta|^2,0,0)
\nonumber\\
&+\frac{1}{2}
              (0, (\cos\theta-\sin\theta)    \text{Im}(\alpha^*\beta),
              (\cos\theta+\sin\theta) \text{Re}(\alpha^*\beta)),              
\end{align}
\end{subequations}
%%%%%%%%%%%%%%%%%%%%%%%%%%%%%%%%%%%%%%%%%%%%%%%
where we have used the relations, $P_a^{\upA_x}=\frac{1}{2}(P_a+U_{xa}), P_a^{\dnA_x}=
\frac{1}{2}(P_a-U_{xa})$.
Physically  $\boldsymbol{P}^{\upA_x}$,$\boldsymbol{P}^{\dnA_x}$
  represents the ppin operator on the  $x$-spin up, $x$-spin down  state
 individually.
 For instance, the electron in the front-layer with $x$ spin-up 
$|\text{f}\upA_x\rangle = \frac{1}{\sqrt{2}}(|\text{f}\upA\rangle + 
|\text{f}\dnA\rangle)$ is the eigenstate of $P^{\upA_x}_z$ with eigenvalue 
 $+\frac{1}{2}$.
 In the C-phase, the ppins exhibit antiferromagnetic  properties
 in the two dimensional spin $y$-$z$ polarized 
 plane, while ppin ferromagnetic properties  
 in the spin $x$ direction.

The ppin N\'{e}el  parameter 
$ \boldsymbol{N}_{ppin}= \boldsymbol{P}^{\upA_x} - \boldsymbol{P}^{\dnA_x}$,
 which is equivalent to $U$-operator $U_{xa}$,
is introduced as another order parameter of the C-phase.
In the spin and ppin phase,
 $\langle\Phi_{S}|\boldsymbol{N}_{ppin}|\Phi_{S}\rangle =
\langle\Phi_{P}|\boldsymbol{N}_{ppin}|\Phi_{P}\rangle =\bold{0}$,
 while in the C-phase, 
%%%%%%%%%%%%%%%%%%%%%%%%%%%%%%%%%%%%%%%%%%%%%%
\begin{align}
&\langle\boldsymbol{N}_{ppin}\rangle_C
\nonumber\\
& =-\frac{1}{2}(0, (\cos\theta-\sin\theta)    \text{Im}(\alpha^*\beta),
              (\cos\theta+\sin\theta) \text{Re}(\alpha^*\beta)).
\end{align}
%%%%%%%%%%%%%%%%%%%%%%%%%%%%%%%%%%%%%%%%%%%%%%%%%%%%
Thus, the ppins are also  ``canted'' in the C-phase. 

Let us examine  the case  $\theta$=0, where  $|v\rangle$ 
 becomes an eigenstate of the operator $P_x$ 
 with the eigenvalue 1.
There are no quantum fluctuations around the $P_x$ axis, where
  the semiclassical interpretation becomes much clear.
The magnitude of  the SU(4) isospin expectation value 
(\ref{SU4C}) becomes unity.
The spin and the ppin  N\'{e}el  order parameters are given as 
%%%%%%%%%%%%%%%%%%%%%%%%%%%%%%%%%%%%%%%%%%%%%%
\begin{subequations}
\begin{align}
&\langle\boldsymbol{N}_{spin}\rangle_C
 =-(\text{Re}(\alpha^*\beta),
             \text{Im}(\alpha^*\beta),0),\\
&\langle\boldsymbol{N}_{ppin}\rangle_C
 =-(0,\text{Im}(\alpha^*\beta),
             \text{Re}(\alpha^*\beta)).
\end{align}
\end{subequations}
%%%%%%%%%%%%%%%%%%%%%%%%%%%%%%%%%%%%%%%%%%%%%%%%%%%%
They are related to each other with 
 the interchange of $x$,$z$ component,
 because in the SU(4) formalism
the spin and  the ppin are treated equivalently.
%$|t\rangle$,$|v\rangle$ are eigenstates of $S_z$,$P_x$.
The interchange of  $x$,$z$ comes from the fact 
 that $|t\rangle$ is an eigenstate of $z$ spin operator $S_z$,
 while $|v\rangle$  is an eigenstate of $x$ ppin operator $P_x$ at $\theta$=0.

\begin{table}
\renewcommand{\arraystretch}{1.5}
\begin{center}
\begin{tabular}{|c|c|c|c|}  
  \hline Phase &  Spin phase & C-phase & Ppin phase\\  
\hline Order Parameter  & $\bbox{S}$ & \ $U_{ab}$ & $\bbox{P}$
\\ \hline       
\end{tabular}
\end{center}
\caption{The order parameters in the three phases.
By including the C-phase, all SU(4) operators appear as  order parameters.} 
\label{order}
\end{table}

\section{SU(4) coherent effects to the C-phase} \label{SecSU(4)Cohe}

\begin{figure}[htbp]
  \begin{center}
   \scalebox{0.7}{\includegraphics{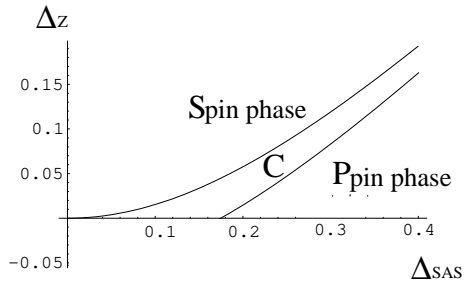}}
  \end{center}
  \caption{Phase diagram in the case $d=\ell_{B}$. 
All energies are in units of 
  $e^2$/$4\pi\epsilon\ell_B$. }
  \label{fig:fig1}
\end{figure}

\begin{figure}[htbp]
  \begin{center}
   \scalebox{0.7}{\includegraphics{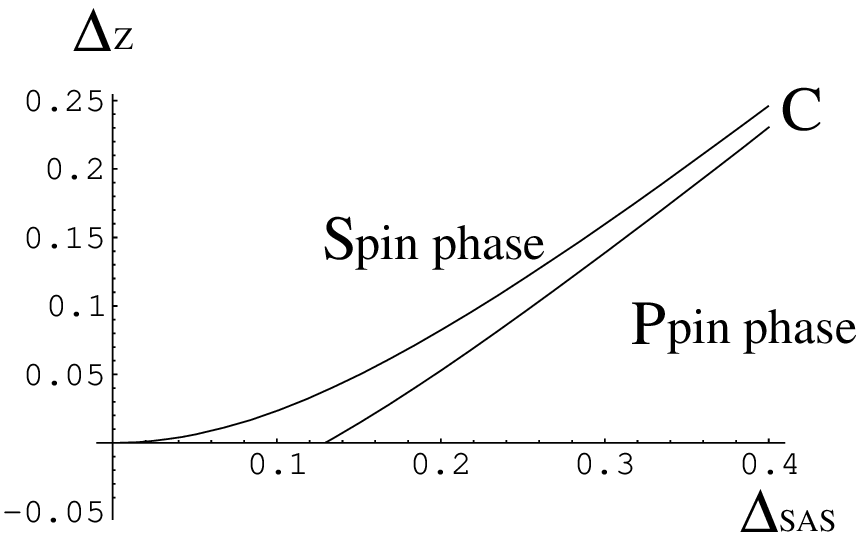}}
  \end{center}
  \caption{Phase diagram  in the case $d=\frac{1}{2}\ell_{B}$. }
  \label{fig:fig2}
\end{figure}

\begin{figure}[htbp]
  \begin{center}
   \scalebox{0.7}{\includegraphics{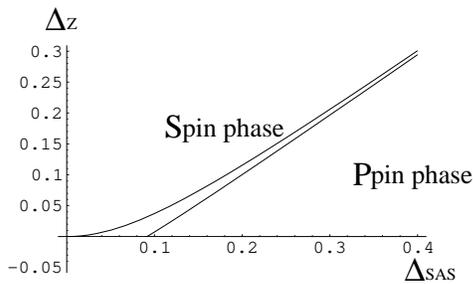}}
  \end{center}
  \caption{Phase diagram in the case $d=\frac{1}{4}\ell_{B}$. }
  \label{fig:fig3}
\end{figure}

 With the use of (\ref{tmin}), we investigate SU(4) coherent effects to the C-phase  
by varying the interlayer  separation $d$
 [Fig. \ref{fig:fig1}, \ref{fig:fig2}, \ref{fig:fig3}].

%%%%%%%%%%%%%%%%%%%%%%%%%%%% 

%%%%%%%%%%%%%%%%%%%%%%%%%%%% 
%%%%%%%%%%%%%%%%%%%%%%%%%%%% 

%%%%%%%%%%%%%%%%%%%%%%%%%%%% 
%%%%%%%%%%%%%%%%%%%%%%%%%%%%

It is obvious from these diagrams,  
when two layers are close,  the C-phase region becomes narrow.
For instance, in [Fig.\ref{fig:fig3}] which is in 
the case $d=\frac{1}{4}\ell_{B}$,
 the  C-phase almost vanishes being eaten up by  the ppin phase.   
The SU(4) coherence disturbs the realization of the  C-phase.

As we have discussed in Section\ref{SecGrounState},
in the SU(4) limit
   the C-phase disappears and 
the spin phase suddenly changes to the ppin phase.
% [Fig.\ref{d0}]. 
However  when $d$ becomes larger, 
the SU(4)-noninvariant exchange term cannot be neglected.
The exchange energies of the spin-sector and the ppin-sector %six states 
  are no longer 
degenerate.
When we begin from the $v$-boson condensation state (the ppin phase),
 accompanied with  the increase of the interlayer separation $d$,
 the exchange energies of $v$-bosons (\ref{TrueEnergP})
will increase.
To lower the energy loss made by  
the SU(4)-noninvariant exchange interaction, 
the  $v$-bosons and $t$-bosons  cooperate 
and condense simultaneously, because the exchange energy of $t$-boson 
 is always lower than that of $v$-boson.
By involving  $t$-bosons,  %if the linear combination is taken,
 the exchange energy  actually decreases, while  the direct energy increases.
Due to the competition of  these two effects,  the system  
settles in another energy minimum, which is  C-phase.

 As we have seen in Subsection \ref{C-phaseene},
 the $U$-operators  in (\ref{U++etc}),(\ref{U+zetc})
 change the $v$-states  to  the $t$-states and vice versa.
 $U$-operators  have  such effects to
 distribute  $t$ and  the $v$ bosons simultaneously.
 However the    $U$-operators (\ref{U++etc}) work  ``contrary''
  to the   $U$-operators in (\ref{U+zetc}), i.e.
 these two kinds of $U$-operators 
   suppress their effects in each other.
 The strength of the  $U$-operators in (\ref{U++etc}) is 
 the ppin stiffness $J_d$, while the strength of the 
  $U$-operators in (\ref{U+zetc}) is  the spin stiffness $J$.
 The difference between the spin and ppin stiffness 
 generates a net effect to condense $t$ and $v$-bosons simultaneously.
 When  two layers are very close, the ppin stiffness $J_d$ are
 nearly equal to the spin 
 stiffness $J$. Hence 
 their effects are almost suppressed.
 Consequently,   the C-phase region is very narrow 
 in the SU(4) coherent region. 

\section{Summary and Discussion}\label{SecSummary}
We have investigated  an origin of the C-phase in the SU(4) context.
 We have seen  both spins and ppins are 
``canted'' in the C-phase.
The $U$-operators become an order parameter of the C-phase.
Using the Schwinger boson  picture, 
we have shown that SU(4) coherence decreases the 
C-phase region.
Microscopically, in the C-phase  
 $t$ and $v$-bosons  condense simultaneously.
 Each $U$-operator has  an effect to induce  simultaneous
 condensation of such bosons.
However, in the SU(4) coherent region,  $U$-operators  compete
 in each other and suppress their effects. 
Consequently the C-phase is unlikely realized in such a region.
\vspace{0.5cm}
\section*{Acknowledgement}
The author would like to acknowledge  Z.F.Ezawa for introducing 
him to this subject and  helpful discussions.
He is very glad to K.Moon for valuable comments. 
This work was supported by grant No. R01-1999-00018 from the
Interdisciplinary Research Program of the KOSEF, the special grant
of Sogang University in 2002 and JSPS fellowship.

%%%%%%%%%%%%%%%%%%%%%%%%%%%%%%%%%%%%%%%%%%%%%%%%%%%%%%%%%%%

\end{document}